\documentclass[twocolumn,aps, showpacs]{revtex4}

\begin{document}

\title{Nonlinear modulation of transverse dust lattice waves
\\ in complex plasma crystals\footnote{Preprint, submitted to \textit{Physics of
Plasmas.}}}
\author{I. Kourakis\footnote{On leave from: U.L.B. - Universit\'e Libre de Bruxelles,
Physique Statistique et Plasmas C. P. 231, Boulevard du Triomphe, B-1050 Brussels,
Belgium;
also:
Facult\'e des Sciences Apliqu\'ees - C.P. 165/81 Physique
G\'en\'erale, Avenue F. D. Roosevelt 49, B-1050 Brussels, Belgium;
\\Electronic address: \texttt{ioannis@tp4.rub.de}} and P. K.
Shukla\footnote{Electronic address: \texttt{ps@tp4.rub.de}}}
\affiliation{Institut f\"ur Theoretische Physik IV,
Fakult\"at f\"ur Physik
und Astronomie, \\
Ruhr--Universit\"at Bochum, D-44780 Bochum, Germany}

\date{Submitted 16 December 2003; revised version 30 January 2003}

\begin{abstract}
The occurrence of the modulational instability (MI) in transverse
dust lattice waves propagating in a one-dimensional dusty plasma
crystal is investigated. The amplitude modulation mechanism, which
is related to the intrinsic nonlinearity of the sheath electric
field, is shown to destabilize the carrier wave under certain
conditions, possibly leading to the formation of localized
envelope excitations. Explicit expressions for the instability
growth rate and threshold are presented and discussed.
\end{abstract}

\maketitle

\paragraph{Introduction.}

Studies of numerous collective processes \cite{psbook} in dust
contaminated plasmas (DP) have been of significant interest in
connection with linear and nonlinear waves that are observed in
laboratory and space plasmas. An issue of particular importance is
the formation of strongly coupled DP crystals by highly charged
dust grains, for instance in the sheath region above a horizontal
negatively biased electrode in experiments \cite{Chu, Thomas,
Melzer, Hayashi}. Low-frequency oscillations occurring in these
mesoscopic dust grain quasi-lattices, in both longitudinal and
transverse directions, have been theoretically predicted
\cite{Melandso, farokhi, vladimirov1, tskhakaya, Wang, Morfill},
and later experimentally observed \cite{psbook, Morfill, Homann,
Ivlev2000, Misawa, Piel}. We note that the observation of the
characteristics of transverse vibrations around a levitated
equilibrium position, where the electric and gravity forces are in
balance, has been suggested as a diagnostic tool, enabling the
determination of the grain charge \cite{Melzer, Melzer2, PS1996}.
Recent generalizations taking into account dust charge variations
\cite{Morfill2}, layer coupling \cite{Vladimirov2} and
two-dimensional crystal anisotropy \cite{anisotropic} are also
worth mentioning.

It is known from solid state physics \cite{solid} that lattice
vibrations are inevitably subject to amplitude modulation due to
intrinsic nonlinearities of the medium. Furthermore, the wave
propagation in crystals are often characterized by the
Benjamin-Feir-type modulational instability (MI), a well-known
mechanism for the energy localization related to the wave propagation
in nonlinear dispersive media. The MI mechanism has been
thoroughly studied in the past, mostly in one-dimensional (1d)
solid state systems, where nonlinearities of the substrate
potential and/or particle coupling may be seen to destabilize
waves and possibly lead to localized excitations (solitary waves)
\cite{Remo, Scott}. In the context of plasma wave theory, this
nonlinear mechanism has been investigated in a variety of contexts
since long ago \cite{redpert, Hasegawa}. In a weakly coupled dusty
(or complex) plasma (DP), in particular, new electrostatic wave
modes arise \cite{Verheest, psbook}, whose modulation has been
studied quite recently \cite{AMS, IKPSDIAW}; instability
conditions were shown to depend strongly on modulation
obliqueness, dust concentration and the ion temperature
\cite{IKPSDIAW}.

In principle, nonlinearity is always present in dusty plasma
vibrations, due to the form of the inter-grain electrostatic
interaction potential, which may be of the Debye-H\"uckel--type \cite{Melandso,
tskhakaya} or else \cite{Ignatov, IKPSPLA}. Furthermore, the
electric potential dominating oscillations in the transverse
direction, yet often thought to be practically parabolic near the
levitated equilibrium position \cite{Tomme}, is intrinsically
anharmonic \cite{tskhakaya2}, as suggested by experimental results
\cite{Zafiu, Ivlev2000}. Despite this evidence, knowledge of
nonlinear mechanisms related to low-frequency DP lattice modes is
still in a preliminary stage. Small amplitude localized
longitudinal excitations (described by a Korteweg-de Vries
equation) were considered in Refs. \cite{Melandso, PS2003}
based on which longitudinal dust lattice wave (LDLW)
amplitude modulation was considered in Ref. \cite{AMS2}.
However, to the best of our knowledge, no study has been carried
out, from first principles, of the amplitude modulation of
transverse dust lattice waves (TDLWs) because of the sheath
electric field nonlinearity. This Letter aims in making a first
analytical step towards the study of DP crystals in this
framework.

About three years ago, Misawa {\it et al.} \cite{Misawa} reported
the observations of vertical nonlinear oscillations of a dust
grain in a plasma sheath, and interpreted their results in terms
of a position-dependent delayed grain charging effect. Zafiu
\textit{et al.} \cite{Zafiu} studied vertical dust-grain
oscillations in a rf-discharge plasma, and succeeded in pointing
out their strongly nonlinear behavior due to the sheath potential
anharmonicity. Recently, Ivlev {\it et al.} \cite{Ivlev2003}
investigated the nonlinear coupling between high-frequency
transverse (vertical) dust lattice oscillations (TDLOs) and slow
longitudinal dust lattice vibrations (LDLVs). Within the framework
of a slowly varying envelope approximation, they derived a pair of
equations for the modulated TDLOs and  driven (by the
ponderomotive force of the latter) slow LDLVs. The coupled system
of equations admit envelope soliton solutions. Finally,
compressional pulses were studied, both theoretically and
experimentally, in Ref. \cite{Nosenko}; however, these excitations
are related to longitudinal grain motion.

In this Brief Communication, we consider the amplitude modulation
of transverse dust lattice waves (TDLWs), taking into account
self-interaction nonlinearities associated with harmonic
generation, involving the intrinsic nonlinearity of the sheath
electric potential. Thus, the underlying physics of our nonlinear
process is entirely different from those considered in Refs.
\cite{Misawa, Zafiu, Ivlev2003, Nosenko}. By adopting the
reductive perturbation technique, we derive a cubic nonlinear
Schr\"odinger equation for the modulated TDLWs. It is shown that
the latter may be modulationally unstable depending on the plasma
parameters, and that they can propagate in the form of envelope
localized excitations due to a balance between nonlinearity and
dispersion. Explicit forms of localized excitations are presented.

\paragraph{Equation of motion.}

Let us  consider TDLWs (vertical, off--plane) propagating in the
one-dimensional (1d) DP crystal. The dust grain charge $q$ and the
mass $M$ are assumed to be constant.  DP crystals have been shown
to support low-frequency optical-mode-like oscillations in both
transverse and longitudinal directions \cite{psbook, Melandso,
farokhi, vladimirov1, tskhakaya, Wang}. Focusing on the former and
summarizing previous results, let us recall that transverse motion
of a charged dust grain (mass $M$, charge $q$, both assumed
constant for simplicity) in a DP crystal (lattice constant $r_0$)
obeys an equation of the form
\begin{equation}
M \, \frac{d^2 \delta z_n}{dt^2} = M \, \omega_0^2 \, (2 \,\delta
z_n - \,\delta z_{n-1} - \,\delta z_{n+1}) + F_e - Mg \, ,
\label{eqmotion}
\end{equation}
where $\delta z_n = z_n - z_0$ denotes the small displacement of
the $n-$th grain around the equilibrium position $z_0$, in the
transverse ($z-$) direction, propagating in the longitudinal
($x-$) direction. Assuming that the neighboring dust grains
(situated at a distance $x = |x_i - x_j|$) interact via an
electrostatic potential  $\Phi(x)$, we obtain the DP oscillation
`eigenfrequency' $\omega_0^2\,  = - \frac{q}{M r_0} \biggl.
\frac{\partial \Phi(x)}{\partial x}\biggr|_{x=r_0}$, e.g. in the
case of a Debye-H\"uckel potential: \( \Phi(x) = ({q}/{x}) \,e^{-
{x/\lambda_D}}\),
\begin{equation}
\omega_0^2\,  = \frac{q^2}{M r_0^3} \, \biggl(1 +
\frac{r_0}{\lambda_D}\biggr) \,e^{-r_0/\lambda_D} \, ,
\label{Debye-frequency}
\end{equation}
where $\lambda_D$ denotes the effective DP Debye radius
\cite{psbook}. The force $F_e = q\, E(z)$ is due to the electric
field $E(z) = - \partial V(z)/\partial z$; the potential $V(z)$ is
obtained by solving Poisson's equation, taking into account the
sheath potential and also (in a more sophisticated description)
the wake potential generated by supersonic ion flows towards the
electrode \cite{PS1996}. The potential $V(z)$ thus obtained,
actually a {\em nonlinear} function of $z$, can be developed
around the equilibrium position $z_0$ as
\begin{eqnarray}
V(z) \approx V(z_0) + V_{(1)} \, \delta z + \frac{1}{2}\, V_{(2)}
\, (\delta z)^2 + \frac{1}{6} \,V_{(3)} \, (\delta z)^3 \nonumber
\\ + {\cal O}[(\delta z)^4] \,; \label{Vseries}
\end{eqnarray}
obviously $V_{(j)} \equiv \biggl. \frac{\partial^j V(z)}{\partial
z^j}\biggr|_{z=z_0}$; the electric force therefore reads
\[
F_e(z) \approx F_e(z_0) + \gamma_{(1)} \, \delta z + \gamma_{(2)}
\, (\delta z)^2 + \gamma_{(3)} \, (\delta z)^3 + {\cal O}[(\delta
z)^4] \, ,
\]
where all coefficients are defined via the derivatives of
$V(z)$, i.e. $\gamma_{(j)} \equiv - q\, \frac{1}{j!} V_{(j+1)}$.
The zeroth-order term balances gravity at (and actually defines
the value of) $z_0$, viz. \( F_e(z_0) - M g = 0 \), while $-
\gamma_{(1)} = q \, V_{(2)}= \gamma \equiv M \, \omega_g^2 $ is
the effective width of the potential well; the value of the gap
frequency $\omega_g = \omega(k=0)$ may either be evaluated from
first theoretical principles \cite{vladimirov1} or determined
experimentally \cite{Ivlev2000}, and is typically of the order of
$\omega_g/2 \pi \approx 20\, Hz$ in laboratory experiments.
Collisions with neutrals and dust charge dynamics are omitted, at
a first step, in this simplified model.

Retaining only the linear contribution and considering phonons of the
type, $x_n = A_n \,exp[i \,(k n r_0 - \omega t)] + c.c.$, we obtain an
optical-mode-like dispersion relation
\begin{equation}
\omega^2\,  = \omega_g^2\, - 4 \omega_0^2\, \sin^2 \biggl( \frac{k r_0}{2}
\biggr)
\, , \label{dispersion-discrete}
\end{equation}
where $\omega$ and $k = 2 \pi/\lambda$ denote, respectively, the
wave frequency and the wavenumber. We will not go into further
details concerning the linear regime, since it is sufficiently
covered in Refs. \cite{Melandso, farokhi, vladimirov1, tskhakaya,
Wang}. Let us now see what happens if the {\em nonlinear} terms
are retained.

\paragraph{Derivation of a Nonlinear Schr\"{o}dinger Equation.}

For analytical tractability, we shall limit ourselves to a
quasi--continuum limit, by considering an amplitude which varies
over a scale $L$ which is significantly larger than the
inter-grain distance $r_0$ (i.e. $L/r_0 \ll 1$). Equation
(\ref{eqmotion}) takes the form
\begin{equation}
\frac{d^2 u_{n}}{dt^2} + \,  c_0^2 \,(2 \,\delta u_{n} - \,u_{n-1}
- \,u_{n+1}) + \omega_g^2 \, u_{n} + \alpha \, u_{n}^2 + \beta \,
u_{n}^3 = 0 \, , \label{NL-eq}
\end{equation}
where we set $\delta z \equiv u(x, t)$ for simplicity; $c_0 =
\omega_0 \, r_0$ is a characteristic propagation speed related to
the interaction [e.g. Debye--type, see (\ref{Debye-frequency})]
potential; the nonlinearity coefficients $\alpha$, $\beta$ are
related to the anharmonicity of the electric potential, viz.
\begin{equation}
\alpha = -
\frac{\gamma_{(2)}}{M} \equiv \frac{q V_{(3)}}{2 M}\, , \qquad
\beta = - \frac{\gamma_{(3)}}{M} \equiv \frac{q V_{(4)}}{6 M} \, .
\end{equation}
Remember that inter-grain interactions are {\em repulsive},
hence the difference in structure from the nonlinear Klein-Gordon
equation used to describe one-dimensional oscillator chains. `Phonons'
in this chain are stable only in the presence of the electric field (i.e.
for $\gamma \ne 0 $).

We now proceed by considering small-amplitude oscillations of the form
\[u =
\epsilon \, u_{1} + \epsilon^2 \, {u_{2}}^2 + \,...\] at each
lattice site. Introducing multiple scales in time and space, i.e.
$X_n = \epsilon^n \, x$, $T_n = \epsilon^n \, t$ \ (n = 0, 1, 2,
...), we develop the derivatives in Eq. (\ref{NL-eq}) in powers of
the smallness parameter $\epsilon$ and then collect terms arising
in successive orders. The equation thus obtained in each order can
be solved and substituted to the subsequent order, and so forth.
This reductive perturbation technique is a standard procedure for
the study of the nonlinear wave propagation (e.g. in
hydrodynamics, in nonlinear optics, etc.) often used in the
description of localized pulse propagation, prediction of
instabilities, etc.  \cite{Remo, Scott}. This procedure leads to a
solution of the form
\begin{eqnarray}
u(x, t) = \epsilon \,\biggl[A \,e^{i \,(k x - \omega t)} +
c.c.\biggr] \qquad \qquad \qquad \qquad  \nonumber \\ + \,
\epsilon^2 \,\alpha\, \biggl[ - \frac{2 \, |A|^2}{\omega_g^2} \, +
\frac{A^2}{3 \omega_g^2} \,e^{2 i \,(k x - \omega t)} + c.c.
\biggr] \, + {\cal O}(\epsilon^3) \, ,
\end{eqnarray}
where \textsl{c.c.} denotes the complex conjugate; recall that
$\omega$ obeys the dispersion relation
(\ref{dispersion-discrete}).

The slowly-varying amplitude $A = A(X_1 - v_g \, T_1)$ moves at
the (negative) group velocity $v_g = d\omega/dk = - \omega_0^2 r_0
\,\sin (k r_0/\omega)$ in the direction opposite to the phase
velocity; this {\em{backward}} wave has been observed
experimentally: see the discussion in Ref. \cite{Misawa}. The
amplitude $A$ obeys a {\em Nonlinear Schr\"{o}dinger Equation}
(NLSE) of the form
\begin{equation}
i\, \frac{\partial A}{\partial T} + P\, \frac{\partial^2
A}{\partial X^2} + Q \, |A|^2\,A = 0 \, , \label{NLSE}
\end{equation}
where the `slow' variables $\{ X, T \}$ are $\{ X_1 - v_g \, T_1,
T_2 \}$, respectively. The {\em dispersion coefficient} $P$, which
is related to the curvature of the phonon dispersion relation
(\ref{dispersion-discrete}) as $P = \,({d^2 \omega}/{d k^2})/2$,
reads
\begin{equation}
P =  - \frac{c_0^2 \omega_0^2}{4 \omega^3}\biggl[ 2 \,
\biggl(\frac{\omega_g^2}{\omega_0^2} - 2 \biggr) \, \cos (k r_0)\, +
\cos (2 k r_0) \, + 3 \biggr] \, , \label{Pcoeff}
\end{equation}
and the {\em nonlinearity coefficient}
\begin{equation}
Q = \frac{1}{2 \omega} \biggl( \frac{10 \alpha^2}{3 \omega_g^2} -
3\, \beta \biggl) \, = \,
\frac{1}{2 M \omega} \biggl[ - \frac{10 \gamma_{(2)}^2}{3 \gamma_{(1)}} +
3\, \gamma_{(3)} \biggl]
\label{Qcoeff}
\end{equation}
is related to the electric field nonlinearity considered above.
Notice that the sign of $P$ depends on the ratio
$\omega_g/\omega_0 \equiv \lambda \, > 0$, which appears naturally
as an order parameter; see for instance in
(\ref{dispersion-discrete}) that $\lambda \ge 2$ is a stability
criterion (necessary for $\omega$ to be real in the whole range of
the first Brillouin zone). For long wavelength values, $P \approx
- ({c_0^2 \omega_g^2})/({2 \omega^3}) \, < 0$, given the parabolic
form of $\omega(k)$ close to $k=0$ (continuum case).
In the general (discrete) case, we see that the
coefficient $P$ becomes positive at some critical value of $k$,
say $k_{cr}$, inside the first Brillouin zone. Some simple algebra
shows that the {\sl zero-dispersion point} $k_{cr}$ satisfies the relation:
$\cos (k_{cr} r_0) = (2 - \lambda^2 + \lambda \sqrt{\lambda^2 -
4})/2$, for $\lambda > 2$ ($0 < k_{cr} r_0 < \pi$); otherwise, for
$\lambda < 2$, $P$ remains negative everywhere.

\paragraph{Modulational instability.}

In a generic manner, a modulated wave whose amplitude obeys the
NLS equation (\ref{NLSE}), is stable (unstable) to perturbations if
the product \( P Q \) is negative (positive).
To see this, one may first check that the NLSE accepts the
monochromatic solution (Stokes' wave)
\( A(X, T) = A_0\, e^{i Q
|A_0|^2 T} \, + \, c.c. \) The standard (linear) stability
analysis then shows that a linear modulation with the frequency
$\Omega$ and the wavenumber $\kappa$ obeys the dispersion relation
\begin{equation}
\Omega^2(\kappa) = P^2\, \kappa^2\, \biggl( \kappa^2\, - 2
\frac{Q}{P}\,|A_0|^2 \biggr) \, ,
\label{pert-disp}
\end{equation}
which exhibits a purely growing mode for
\(\kappa \geq \kappa_{cr} =
({Q}/{P})^{1/2}\,|A_0|\). The growth rate attains a maximum value of
\( \gamma_{max} =
{Q}\,|A_0|^2 \). This mechanism is known as
the {\sl Benjamin-Feir instability} \cite{Remo}.
For $P Q < 0$, the wave is modulationally stable,
as evident from (\ref{pert-disp}).

One now needs to deduce the sign of Q, given by (\ref{Qcoeff}), in
order to determine the stability profile of the TDL oscillations.
In fact, given the above definitions of the parameters $\omega$,
$\alpha$, $\beta$, one easily finds that $Q$ is related to the
(derivatives of the) electric potential $V(z)$ via
\begin{equation}
Q = \frac{q}{4 M \omega} \biggl[ \frac{5 V_{(3)}^2}{3 V_{(2)}} -
V_{(4)} \biggl] \, = \, \frac{\omega_g^2}{4 \omega} \biggl[
\frac{5 V_{(3)}^2}{3 V_{(2)}^2} - \frac{V_{(4)}}{V_{(2)}}
\biggl] \, .
\label{Qcoeff-reduced}
\end{equation}
The exact form of the potential $V(z)$ may be obtained from
\textsl{ab initio} calculations or by experimental data fitting.
For instance, in Ref. \cite{Ivlev2000}, the dust grain potential
energy ${\mathcal{U}}(z) = q\, V(z)$ was reconstructed from
experimental data as
\begin{eqnarray}
{\mathcal{U}}(z) \simeq M \omega_0^2\, \biggl[ - \, 0.9 \,\delta z
\,+\, \frac{1}{2} \,(\delta z)^2 \,- \,\frac{1}{3} \, 0.5
\,(\delta z)^3 \, \nonumber \\ +\,\frac{1}{4} \, 0.07 \,(\delta
z)^4 \,+\,... \, \biggr] \, ,
\end{eqnarray}
which is  Eq. (9) in Ref. \cite{Ivlev2000}; upon simple inspection
from (\ref{Vseries}), we obtain \( V_{(3)}/V_{(2)} = -1\, , \quad
V_{(4)}/V_{(2)} = 0.42 \), so the value of $Q$ is positive, as may
be checked from (\ref{Qcoeff-reduced}). Therefore, the transverse
oscillation considered in Ref. \cite{Ivlev2000} would propagate as
a \emph{stable} wave, for large wavelength values $\lambda$.
However, for shorter wavelengths, the coefficient $P = \,
\omega''(k)/2$ - as defined in (\ref{dispersion-discrete}) - may
become positive (and so will the product $P Q$, in this case); the
TDL wave may thus be potentially unstable. These results may
\textsl{a priori} be checked experimentally.

\paragraph{Localized excitations.}

A final comment concerns the possibility of the existence of localized
excitations related to transverse dust-lattice waves. It is known
that the NLSE (\ref{NLSE}) supports pulse-shaped localized
solutions (envelope solitons) of the {\sl bright} ($P Q > 0$) or
{\sl dark/grey} ($P Q < 0$) type \cite{Fedele, Fedele2}.
The former (\textsl{continuum breathers}) are
\begin{eqnarray}
A = (2 D/P Q)^{1/2} \, sech \bigl[ (2 D/P Q)^{1/2}\, (X - v_e \,
T)\bigr]\, \nonumber \\
\times \, \exp\bigl[ i \, v_e \, (X - v_c \, T)/ 2 P\bigr] \, + \,
c.c. \, ,\label{breather}
\end{eqnarray}
where $v_e$ ($v_c$) is the envelope (carrier) velocity and $D =
(v_e^2 - 2 \,v_e \,v_c)/(4 P^2)$; they may occur and propagate in
the lattice if a sufficiently short wavelength is chosen, so that
the product $P Q$ is positive. We note that the pulse width $L$ and
the amplitude $\rho$ satisfy $L \rho \sim (|P/Q|)^{1/2} = const.$

For $P Q < 0$, we have the {\em grey} envelope soliton
\cite{Fedele}
\begin{equation}
A \, = \, \rho_1 \, \{ 1 - a^2 \, sech^2\{[X - (v_e \, + 2 \alpha)
T]/L_1\}\}^{1/2} \, \exp ( i\, \sigma ) \, , \label{greysoliton}
\end{equation}
where $\sigma =  \sigma(X, T)$ is a nonlinear phase correction to
be determined. This excitation represents a localized region of
negative wave density (a \emph{void}), with finite amplitude $(1 -
a) \rho_1$ at $X = 0$; \, \, $0 \le a \le 1$). Again, the pulse
width $L_1 = (|P/Q|)^{1/2}/(a \rho_1)$ is inversely proportional
to the amplitude $\rho_1$. Notice the (dimensionless) parameter
$a$, which regulates the depth of the excitation. For $a = 1$, one
obtains a {\em dark} envelope soliton, which describes a localized
density \textit{hole}, characterized by a vanishing amplitude at
$\zeta = 0$. The latter excitations (of grey/dark type), yet
apparently privileged in the continuum limit (where $P Q < 0$),
are rather physically irrelevant in our (infinite chain) model,
since they correspond to an infinite energy stored in the lattice.
Nevertheless, their existence locally in a finite--sized chain may
be considered (and possibly confirmed) either numerically or
experimentally.

\textsl{In conclusion}, we have shown that the modulational
instability is, in principle, possible for transverse DP lattice
waves. Long wavelength modes seem to ensure wave stability, while
shorter wavelength modes may be modulationally unstable. The
existence of localized excitations and the occurrence of the
modulational instability rely on the same criterion, which needs
to be thoroughly examined for a given exact form of the sheath
electric potential. These results may be investigated and will
hopefully be confirmed by appropriate experiments. Of course, for a
more complete description, one has to take into account certain
factors that are ignored in this simple model, viz. collisions with neutral
particles, dust charge variations and, eventually, transverse to
longitudinal mode coupling. The effect of inter-layer coupling may
also be investigated. Work in this direction is in progress and
the results will be reported soon.

\medskip

\begin{acknowledgments}
This work was supported by the European Commission (Brussels)
through the Human Potential Research and Training Network via the
project entitled: ``Complex Plasmas: The Science of Laboratory
Colloidal Plasmas and Mesospheric Charged Aerosols'' (Contract No.
HPRN-CT-2000-00140).
\end{acknowledgments}

\end{document}